\documentclass[format=acmsmall, review=false, screen=true]{acmart}
\usepackage{booktabs} 
\usepackage[english]{babel}
\usepackage[ruled]{algorithm2e} 
\usepackage{arabtex}
\usepackage{tabularx}
\usepackage{mathtools}
\usepackage{subfig}
\usepackage{multirow}
\usepackage{arabtex}
\usepackage{utf8}

\SetAlFnt{\small}
\SetAlCapFnt{\small}
\SetAlCapNameFnt{\small}
\SetAlCapHSkip{0pt}
\IncMargin{-\parindent}

\setcopyright{acmlicensed}

\acmDOI{0000001.0000001}

\begin{document}
\markboth{M. Iqbal et al.}{CURE: Collection for Urdu Information Retrieval Evaluation and Ranking}
\title{CURE: Collection for Urdu Information Retrieval Evaluation and Ranking}

\author{Muntaha Iqbal}
\orcid{0000-0002-3698-8263}
\affiliation{%
  \institution{Al-Khawarizmi Institute of Computer Science, UET, Lahore}
  \country{Pakistan}}
\email{muntaha.iqbal@kics.edu.pk}
\author{Kamran Amjad}
\affiliation{%
\institution{Al-Khawarizmi Institute of Computer Science, UET, Lahore}
  \country{Pakistan}}
\email{kamran.amjad@kics.edu.pk}
\author{Bilal Tahir}
\affiliation{%
\institution{Al-Khawarizmi Institute of Computer Science, UET, Lahore}
  \country{Pakistan}}
\email{bilal.tahir@kics.edu.pk}
\author{Muhammad Amir Mehmood}
\affiliation{%
\institution{Al-Khawarizmi Institute of Computer Science, UET, Lahore}
  \country{Pakistan}}
\email{amir.mehmood@kics.edu.pk}


\begin{abstract}
Urdu is a widely spoken language with $163$ million speakers worldwide across the globe. Information Retrieval (IR) for Urdu entails special consideration of research community due to its rich morphological features and a large number of speakers. In general, IR evaluation task is not extensively explored for Urdu. The most important missing element is the availability of a standardized evaluation corpus specific to Urdu. In this research work, we propose and construct a standard test collection of Urdu documents for IR evaluation and named it Collection for Urdu Retrieval Evaluation (CURE). We select 1,096 unique documents against 50 diverse queries from a large collection of 0.5 million crawled documents using two IR models. The purpose of test collection is the evaluation of IR models, ranking algorithms, and different natural language processing techniques. Next, we perform binary relevance judgment on the selected documents. We also built two other language resources for lemmatization and query expansion specific to our test collection. Evaluation of test collection is carried out using four retrieval models as well using the stop-words list, lemmatization, and query expansion. Furthermore, error analysis was performed for each query with different NLP techniques. To the best of our knowledge, this work is the first attempt for preparing a standardized information retrieval evaluation test collection for the Urdu language.

\end{abstract}

%
%



\maketitle
\renewcommand{\shortauthors}{M. Iqbal et al.}

\section{Introduction}\label{Into}

Urdu is the official language of Pakistan and is widely spoken in parts of countries like Bangladesh, United Kingdom, Canada, and India. Urdu has an estimated more than $163$ million speakers worldwide\footnote{\url{https://www.ethnologue.com/language/urd}}. It is amongst the top five most spoken languages of the world~\cite{weber2008top}. Urdu finds its roots in Turkic, Persian, and Arabic languages. Most of Urdu vocabulary is borrowed from Persian, and Arabic languages~\cite{hardie2003developing}. 

Information retrieval is the process of storing and retrieving information for a given query~\cite{madankar2016information}. Nowadays, a significant amount of content is available online in many regional languages, and users prefer to search for content in their native languages. Despite a large number of Urdu speakers around the world and availability of Urdu content online, the task of Information Retrieval (IR) in Urdu has not been explored that much as compared to other regional and global languages. Standard test collection is an essential requirement for the evaluation of IR. Test collections are available for many European\footnote{\url{http://trec.nist.gov/data.html}} \footnote{\url{http://clef.isti.cnr.it/}} , Indian\footnote{\url{http://fire.irsi.res.in/fire/2018/home}}, and Asian languages~\cite{aleahmad2009hamshahri,esmaili2014towards}. However, there is the scarcity of such test collections, especially for Urdu. In~\cite{riaz2008baseline}, test collection was developed for evaluation of Urdu IR. This test collection consists of fewer documents and is not available for research purposes. Furthermore, this test collection is not prepared according to the Text Retrieval Conference (TREC) standards. In this work, we focus on developing a comprehensive test collection to evaluate IR for the Urdu language. 

Our major contributions are as follows:
\begin{itemize}

\item \textbf{Test collection creation:}
For the development of test collection, the first corpus was built using crawled documents from Urdu websites. Second, queries are formulated using information needs from different topic categories. For each query, two retrieval models VSM and BM25 are used to retrieve top $20$ documents from crawled corpus. An initial set of $2,000$ documents obtained against $50$ queries was reduced to $1,096$ unique documents.  For all documents, binary relevance judgment was provided. CURE is freely available for the research community online.

\item \textbf{Evaluation of IR models:}
For the evaluation, all retrieved documents were compared against relevance judgments. Three evaluation measures, i.e., Precision, Recall and Mean Average Precision (MAP), was used for the evaluation of four IR models. Experimental results show that more relevant documents are retrieved at top $20$ ranks using BM25 model and Language Model with Jelenik Mercer Smoothing(LM-JMS). MAP for BM25 and LM-JMS is 73\%, 68\% for Language Model Dirichlet Smoothing (LM-DS) and 69\% for VSM.

\item \textbf{Evaluation of IR models with NLP techniques:}
In order to evaluate the impact of different Natural Language Processing (NLP) techniques on the test collection, we select three basic techniques such as stop-words, lemmatization, and query expansion. Already developed stop-words list was used for stop-words evaluation. There were $402$ words which were reduced to $211$. Next, we used the dictionary look-up based approach for lemmatization with total $3000$ word-lemma pair. Finally, we added $2280$ inflectional variants for $642$ words in our list. We evaluated all retrieval models by enabling these NLP techniques. With stop-words removal, results of BM25, VSM, and LM-JMS are improved slightly compared to their baseline models. In case of lemmatization, all four retrieval models, i.e., BM25, LM-JMS, LM-DS, and VSM, showed improved performance slightly. Finally, in case of query expansion BM25, VSM, and LM-JMS showed improved performance. From the error analysis of retrieval models, it was observed that frequency of query terms matters although term frequency is not an indicator of relevance.

\end{itemize}

The rest of the paper is organized as follows: Section~\ref{sec:lit} presents related work on previous test collections. Section~\ref{sec:dataset} introduces CURE test collection and explains the data set generation process and its statistics in detail. Our methodology of information retrieval and language processing techniques are presented in Section~\ref{sec:Methodology}. Section~\ref{sec:results} discusses the results of the evaluation of different IR models on CURE. Section~\ref{sec:conclusion} concludes the paper. 

\section{Related Work}
\label{sec:lit}

The development of resources for evaluation of IR has been the prime focus of the information retrieval research community. Many international platforms are working on the evaluation procedure and have provided the research community with several specialized test collections in different languages. The task of retrieval evaluation is a well-explored one for English and other widely spoken European languages.

Text Retrieval Conference (TREC)\footnote{\url{http://trec.nist.gov/data.html}} is one of the most notable international platforms dedicated to this field \cite{harman1995overview}. This platform contains a test collection for IR for many languages. In addition to the English language, many non-English test collections built for retrieval evaluation of regional languages are also made available through this platform, i.e., German, French, Italian, Arabic, Spanish and, Chinese. Other notable platforms are, NII Test Collections for IR systems (NTCIR)\footnote{\url{http://research.nii.ac.jp/ntcir/outline/prop-en.html}} for retrieval evaluation of cross-lingual and East Asian language text.Cross-Language Evaluation Forum (CLEF)\footnote{\url{http://clef.isti.cnr.it/}} focuses on European languages and Forum for Information Retrieval Evaluation (FIRE)\footnote{\url{http://fire.irsi.res.in/fire/2018/home}} is used for evaluation of IR for most of Indian languages. All these platforms provide test collections for evaluation of monolingual or cross-lingual IR. Several test collections are available for all the platforms mentioned above.

Communications of the ACM (CACM)\footnote{\url{http://www.search-engines-book.com/collections/}} provides a test corpus which consists of $3204$ documents and $52$ queries with relevance judgments for English IR. \textit{CACM} consists of bibliographic data and was developed for the evaluation of the extended VSM~\cite{fox1983characterization}. For evaluation of Slovak IR test collection was created~\cite{hladek2016evaluation}. This collection contains $1097$ documents from an already developed news corpus~\cite{hladek2014slovak} for $80$ queries.

Similarly, for the evaluation of Persian IR, many test collections are available. In~\cite{esmaili2007mahak}, \textit{Mahak} collection was created for Persian IR evaluation. They used $216$ queries and collected $3007$ news documents from Iranian Student's News Agency (ISNA)\footnote{\url{https://www.isna.ir/Main/}}. The most prominent effort for the creation of Persian IR collection is \textit{Hamshahri}~\cite{aleahmad2009hamshahri}. For the selection of documents, they used the pooling method. Documents were collected from news articles from $1996$ to $2002$. It consists of $166,774$ documents and $65$ queries. Documents were selected from $12$ major news categories. For relevance judgment, $17$ Information Technology (IT) students were asked to mark the top $100$ documents if they are relevant to a specific query or not. For each query, they provided narration and description of the query. For the evaluation, they used four versions of Language Models (LM) and $4$ versions of VSM and reported that VSM performed well on their test collection.
\newline
\textit{Pewan}~\cite{esmaili2014towards} is a test collection for evaluation of Kurdish IR and considered as the first standard collection for the Kurdish language. TREC methodology was followed in the construction of Pewan. 
Text documents were collected from news articles of Peyamner\footnote{\url{http://peyamner.net/}} and Voice-of-America\footnote{\url{https://www.dengiamerika.com/}}.  Overall, $115,430$ documents were collected from two news sources. They created $65$ queries initially, and from these queries, they used $45$ to create a pool from $5$ IR systems. In the final collection, they used $22$ queries and excluded those queries which had very large or few relevant documents. They evaluated $5$ retrieval systems on this test collection.

In~\cite{shamshed2010novel}, authors developed the first Bangla collection for IR evaluation. Three major newspapers, Protho-alo\footnote{\url{http://www.prothomalo.com}}, Kaler-Konath\footnote{\url{http://www.kalerkantho.com/}} and Amer-Desh\footnote{\url{http://www.amardeshonline.com}} were used for document collection. Total $69$ documents were selected from these three sources against $14$ queries and $10$ IT students were selected for relevance judgment. Precision and recall measures were used for evaluation and reported average precision and recall of 52\% and 82\%, respectively. A benchmark for Japanese IR was developed and named as \textit{BMIR-J2} \cite{sakai1999bmir}. There are a total of $5,080$ text documents for $60$ queries. 
Test collection is also available for the evaluation of medical Information retrieval, i.e., \textit{Medlar} collection consists of 18 queries, 273 documents, and selective judgment by query authors. Some other data sets are extended Medlar, \textit{Ophthalmology1}, and \textit{Ophthalmology2}~\cite{salton1973recent}.
\newline
All the platforms mentioned above do not provide any such corpus for the Urdu language that can be employed for Urdu specific IR performance evaluation. This can be attributed to the fact that there has not been much research work in the field of IR for the Urdu language. One such foundational work is described in~\cite{riaz2008baseline}. They used the Backer-Riaz~\cite{becker2002study} corpus, which consists of $7,000$ Urdu news documents. From these $7,000$ documents, a baseline test collection was developed. They selected $200$ documents for their dataset and created a relevance judgment for each document. $4$ queries were created for baseline evaluation. The work done is not comprehensive, and the test collection is not available for public access. Besides this work, there is no mentionable work that is related to this task. In this context, our primary intention was to develop a test collection for evaluation of Urdu IR.
Table~\ref{tab:1st} summarizes characteristics of existing test collections.
\begin{table}
\caption{Summary of existing test collections and CURE}
\label{tab:1st}
\begin{minipage}{\columnwidth}
\begin{center}
\begin{tabular}{llllll}
 \toprule
\textbf{Collection Name} & \textbf{Documents} &\textbf{Relevant Documents} & \textbf{Queries} &\textbf{Domain} &\textbf{Available}\\
\bottomrule
TREC-1	& 741,856	& 277/Query	& 50	& English	& Yes\\
TREC-2	& 741,856	& 210/Query	& 50	& English	& Yes\\
TREC-3	& 741,856	& 196/Query	& 50	& English	& Yes\\
TREC-4	& 741,856	& 130/Query	& 50	& English	& Yes\\
CACM	& 3204	& 796	& 52	& English	& Yes\\
Slovak Collection	& 3980	& 1097	& 80	& Slovak	& No\\
Mahak	& 3007	& N/A	& 216	& Persian	& No\\
Hamshari	& 166,774	& 2352	& 65	& Persian	& No\\
Pewan	& 115,430	& N/A	& 22	& Kurdish	& No\\
Bangla Collection	& 69	& 43	& 14	& Banglali	& No\\
BMIR-J2	& 5080	& 28/query	& 60	& Japanese	& No\\
Medlar	& 273	& N/A	& 18	& Medical	& No\\
Extended Medler	& 450	& N/A	& 29	& Medical	& No\\
Ophthalmology 1	& 852	& N/A	& 29	& Medical	& No\\
Ophthalmology 2	& 852	& N/A	& 17	& Medical	& No\\
Baker-Riaz	& 200	& N/A	& 4	& Urdu	& No\\
CURE	& 500,000	& 1,096	& 50	& Urdu	& Yes\\
 \bottomrule
\end{tabular}
\end{center}
\bigskip\centering
\end{minipage}
\end{table}

\section{CURE Test Collection}
\label{sec:dataset}

This section presents an overview of the data collected in this paper for Urdu IR evaluation. First, we explain the guidelines and the process of query creation. Next, we provide details regarding document collection. Finally, we outline the relevance judgment procedure of these collected documents.

\subsection{Query creation guidelines}
\label{subsec:query1}
For query creation, we acquired assistance from three native speakers of Urdu to specify information need and create queries. We trained our subjects to specify information need and query like reported in~\cite{esmaili2014towards}. Besides, some predefined guidelines were also provided for query creation to subjects~\cite{Paul2013Evaluating}. These guidelines are listed as follows.

\begin{figure}
\begin{center}
\includegraphics[width=10cm,height=8cm,keepaspectratio]{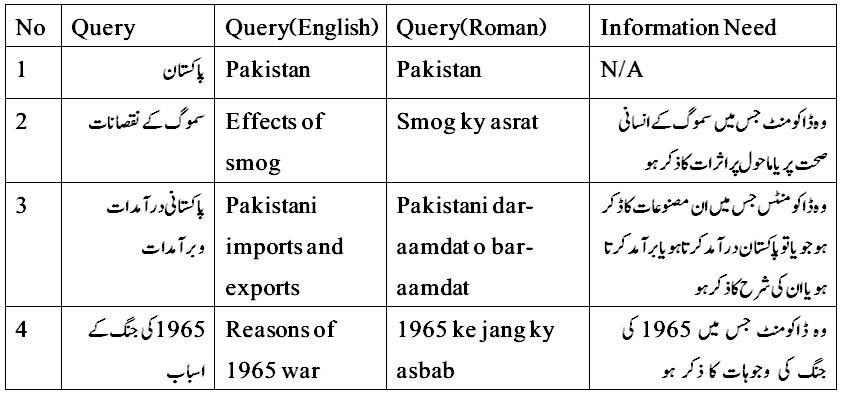}
\caption{Information needs and respective queries}
\label{fig:QIN}
\end{center}
\end{figure}

\begin{itemize} 
\item \textbf{Formulating queries from information needs:} Subjects were first asked to express their information needs regarding a topic. After that, a domain expert provided guidelines to subjects about transforming these information needs to queries. According to guidelines, subjects should identify those words from the information need that represent the main idea and are crucial to retrieve relevant documents. 

Besides, subjects were asked to formulate brief queries. Fig.~\ref{fig:QIN} shows four sample information needs and their respective queries. The description of information needs provided with the query is useful for subjects in the relevance judgment phase~\cite{kinney2008evaluator}. For instance, for the second query in Fig.~\ref{fig:QIN} ``Effects of smog (smog ky asrat)" the user is interested in those documents in which the adverse effects of smog on human beings or environment are discussed.

\item \textbf{Query length:} Length of the query should be higher than $1$ word because the notion of the relevance of a document for a single word query is ambiguous. Example of single word query is shown in the first row of Fig.~\ref{fig:QIN}.

\item \textbf{Total queries:} For standard test collection, there should be a minimum of $50$ queries in the test collection; therefore, each subject was asked to provide at least $20$ queries.

\item \textbf{Query categories:} Queries should be selected from diverse categories for a content-rich test collection. We provided four pairs of information needs and sample queries to our subjects in order to train them. These queries were from history, sports, business, and, health topics.
\end{itemize}

\subsection{Query creation process}
\label{subsec:query}

In the first phase, subjects define the set of the user information needs - the information the user is seeking various topics present in our corpus and prepare a set of corresponding queries. Initially, each subject created $20$ queries to get a pool of $60$ queries. In order to get precise user information needs, we remove all such queries with length more than seven words. 

\begin{figure}
\begin{center}
\includegraphics[width=5.5cm,height=5.5cm,keepaspectratio]{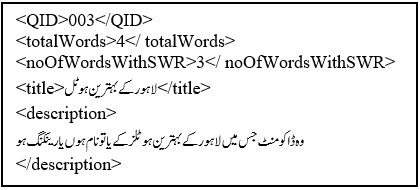}
\caption{Sample query from CURE}
\label{fig:query}
\end{center}
\end{figure}

In assigning a relevance judgment to a document for a query with length higher than $7$ words, subjects were confused, so we discarded all such queries with length higher than $7$ words. After removing these queries, we had $50$ queries in our test collection, which is in accordance with the work is done in~\cite{urbano2016test} where authors suggested a minimum of $50$ queries for standard data collection. Queries in our standard test collection are represented in the standardized format. Fig.~\ref{fig:query} shows an example of a four-word query from CURE. 

\begin{figure}
\begin{center}
\includegraphics[width=7.5cm, height=5.5cm]{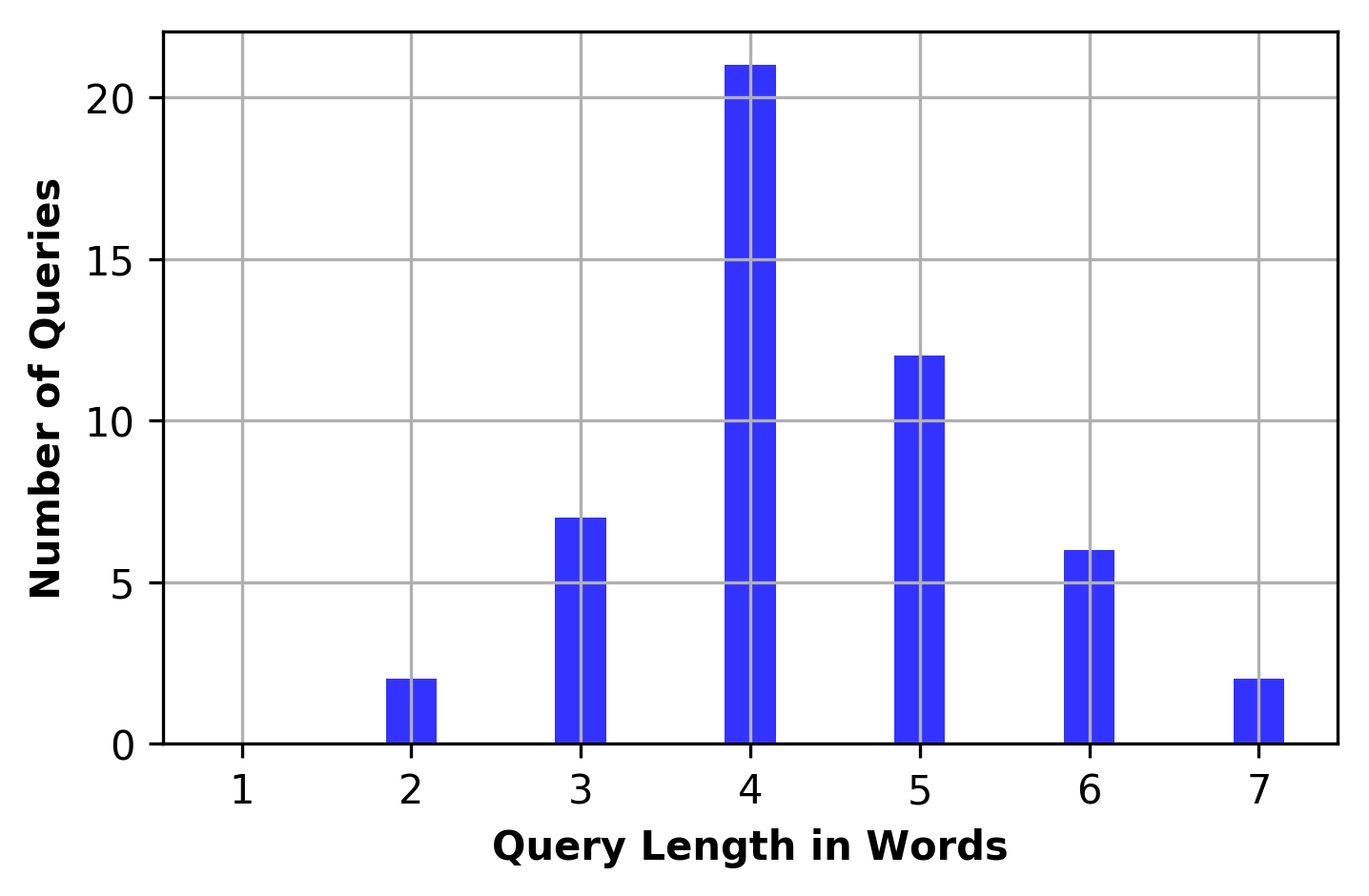}
\caption{Length distributions of queries}
\label{fig:len}
\end{center}
\end{figure}

Every query is assigned five different fields: (i) `QID' - A unique query id, (ii) `totalWords' field shows the total number of words, (iii) `noOfWordsWithSWR' shows the number of words after stop word removal, (iv) `title' field contains the query that was performed, and (v) a `description' field which contains the actual information need based on which that query was formulated. 

\begin{figure}
\begin{center}
\includegraphics[width=7.5cm, height=7.5cm]{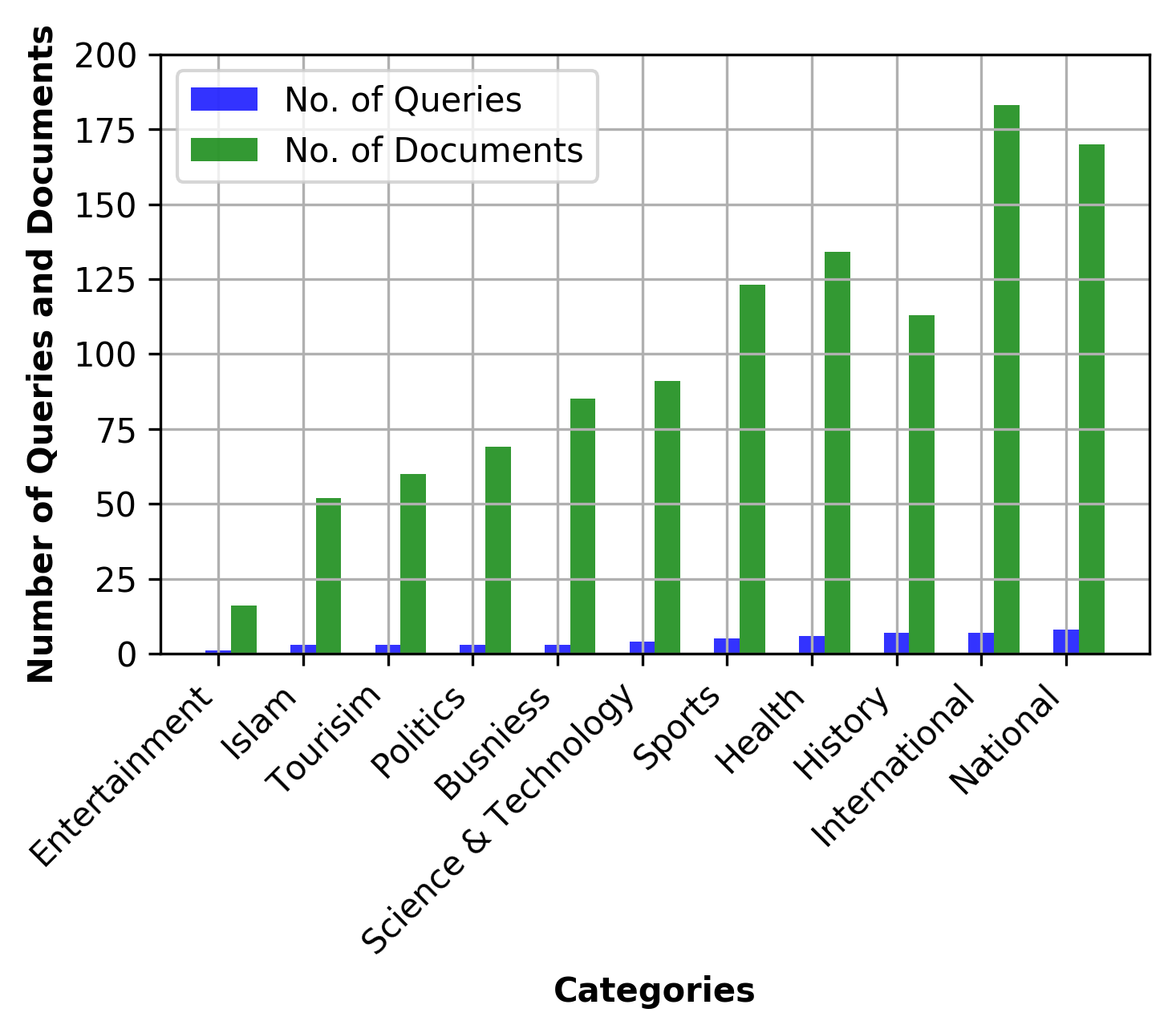}
\caption{Distribution of queries and documents over categories}
\label{fig:cat}
\end{center}
\end{figure}

Next, we present different characteristics of queries. Fig.~\ref{fig:len} shows the length distribution of queries. The average length of $50$ queries in our data set is $4.38$. The maximum length of a query in our data set is $7$, while the minimum length is $2$. We group queries into different topic categories based on different information needs specified by users. Distribution of documents and queries over different topic categories is shown in Fig.~\ref{fig:cat}. We found that 30\% of queries in our test collection belonged to national and international news categories. On average, $21$ documents were retrieved for each query.

\subsection{Document collection}
\label{subsec:Document}
After successful creation of queries, the second step is to select documents based on these queries from a large corpus. To obtain Urdu documents, we employed an open source web-crawler Nutch version (2.3.1), to fetch Urdu web documents present on Urdu websites. Our web-crawler fetched $0.5$ million documents between August 2017 and November 2017. We note that the majority of these documents were crawled from Urdu news websites that frequently post Urdu content. Next, we indexed these documents in \textit{Apache Solr} version (6.2.2) with document id, title, and content.
 
 \begin{figure}
\begin{center}
\includegraphics[width=6.5cm,height=6cm]{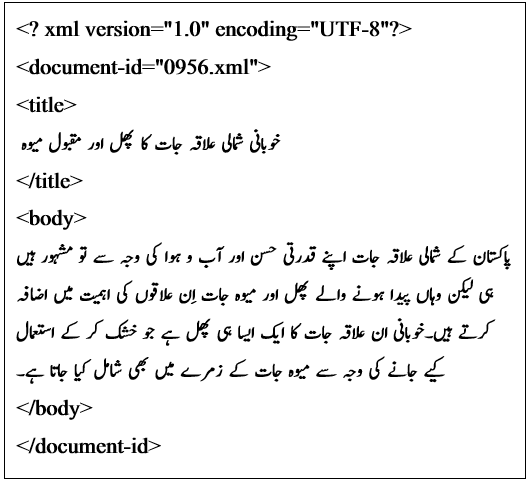}
\caption{Sample document from CURE}
\label{fig:docsample}
\end{center}
\end{figure}

In order to select documents from $0.5$ million corpus, we use the pooling method. Pooling is a standard method in which a subset of `top k' documents is attained using different retrieval models from a corpus~\cite{manning2008introduction}. The top $20$ ranked documents were selected using two retrieval models (VSM, BM25) in Apache Solr6 for the creation of the initial set of documents. After identifying duplicate documents in this subset, we obtained $1096$ documents from an initial pool of $2000$ documents from $254$ domains. We stored our data in a standardized XML format. Each document has: (i) `document\_ID', (ii) `title' field carries the title of the web document, and (iii) `body' field contains the content of that document. Fig.~\ref{fig:docsample} shows sample document from our test collection. The total size of the collection is $6.7$ MB.

\begin{figure}
\begin{center}
\includegraphics[width=7.5cm, height=5.5cm]{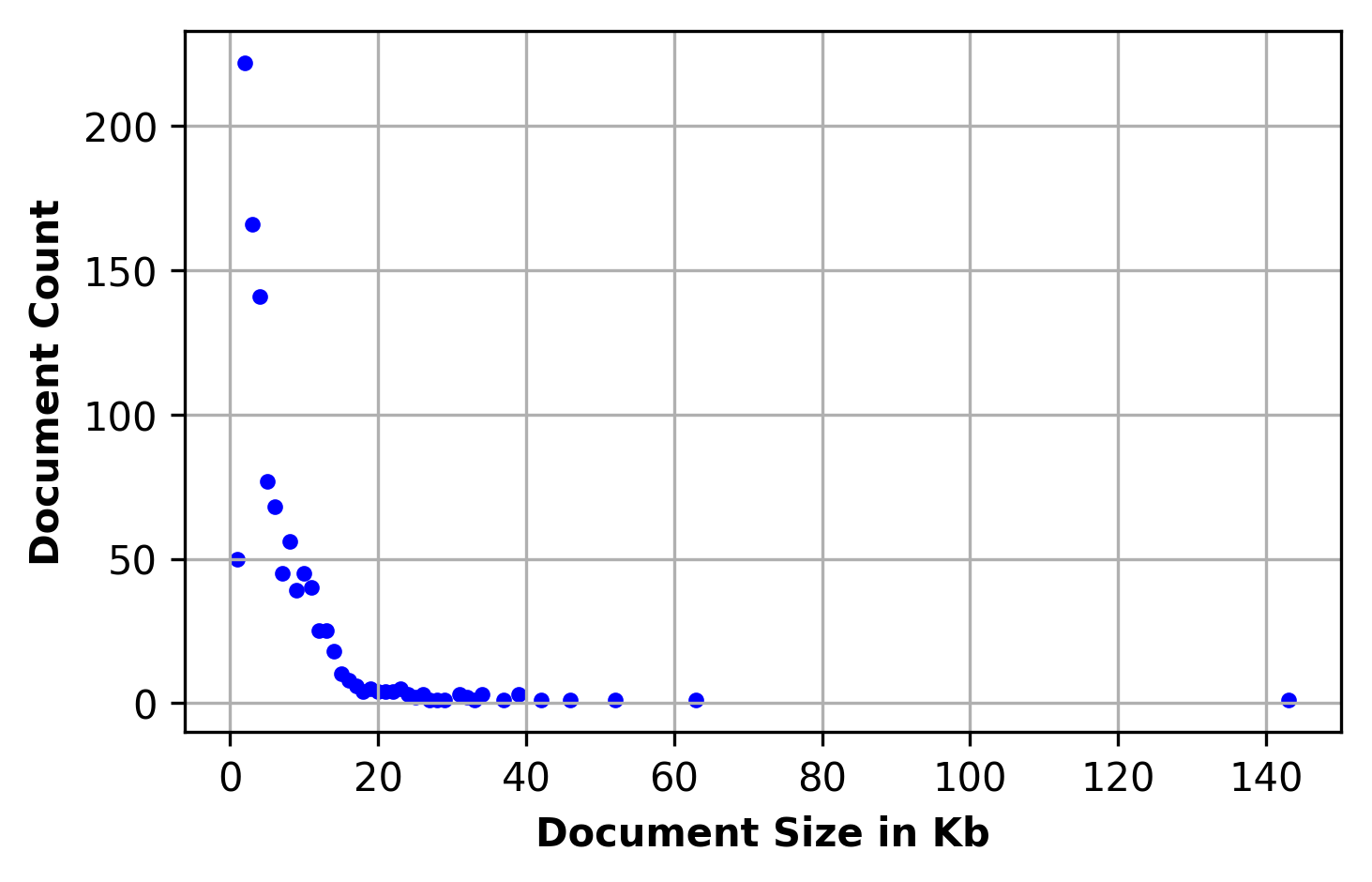}
\caption{Document size distribution}
\label{fig:doclen}
\end{center}
 \end{figure}

 \begin{figure}
 \begin{center}
\includegraphics[width=7.5cm, height=5.5cm]{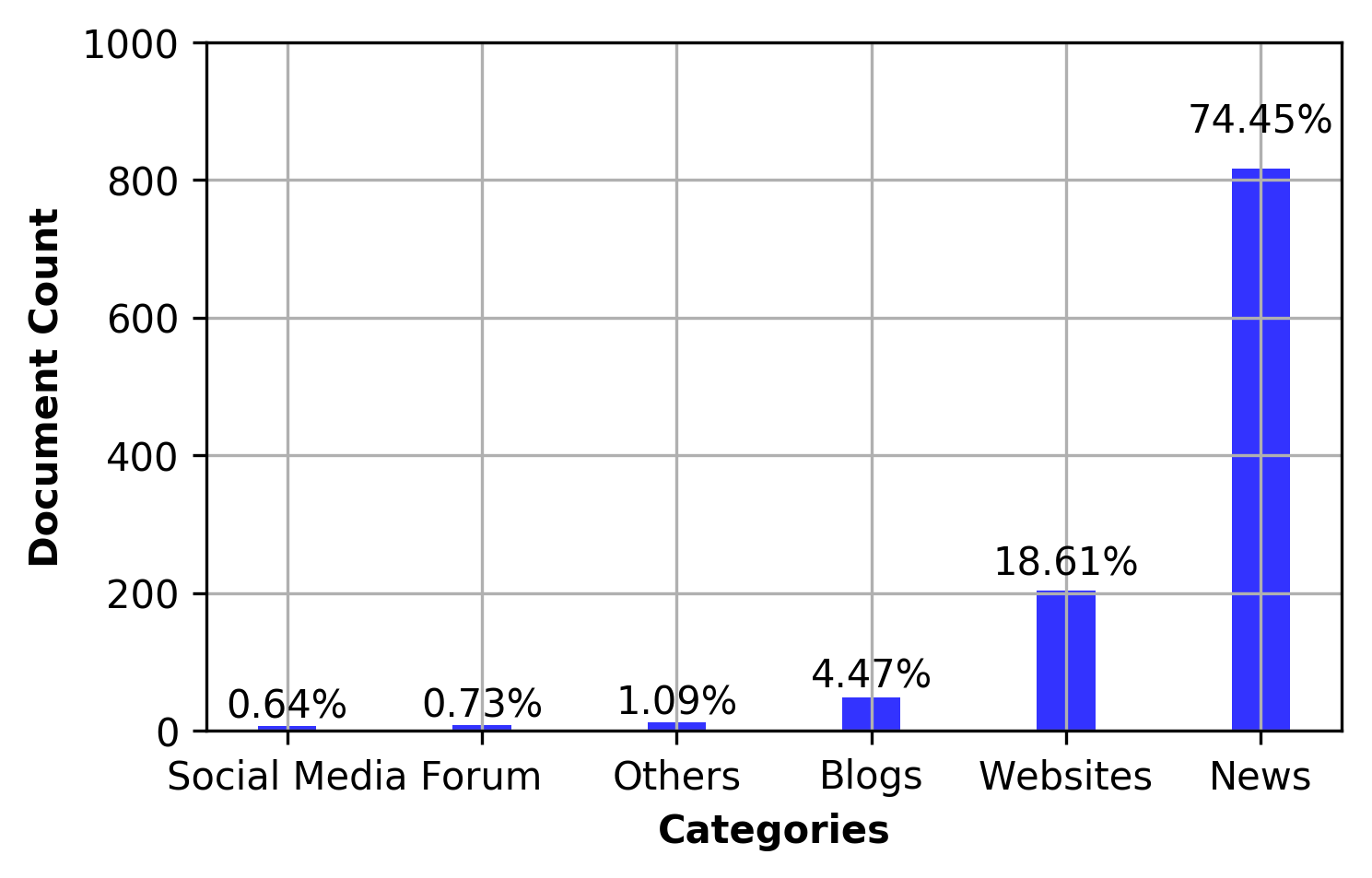}
\caption{Distribution of documents over categories}
\label{fig:Scat}
\end{center}
 \end{figure}

\begin{figure}
\begin{center}
\includegraphics[width=7cm,height=6.5cm]{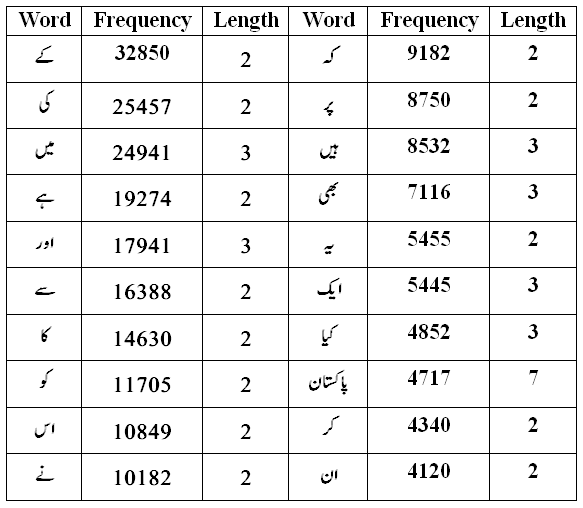}
\caption{Most frequent terms in dataset}
\label{fig:freq}
\end{center}
\end{figure} 

Fig.~\ref{fig:doclen} shows the size of the distribution of documents in the collection of 1096 documents. The size of documents varies from $1$ KB to $143$ KB. More than 80\% of the documents are from $1$ to $10$ KB in size. Fig.~\ref{fig:Scat} shows the distribution of documents based on the source website categories. Almost 75\% of the documents belonged to Urdu news sites. 18\% of the documents belonged to the general website category. 

Besides, documents were also retrieved from blogs, forums, and social media sites. Fig.~\ref{fig:freq} shows the most frequent words in our test collection and their length in characters. Almost all words are stop words except "Pakistan" because in $6$ queries this word is used. Table~\ref{tab:attributes} provides details of different attributes of CURE.

\begin{table}
\caption{Attributes of CURE}
\label{tab:attributes}
\begin{minipage}{\columnwidth}
\begin{center}
\begin{tabular}{ll}
 \toprule
 \textbf{Attributes} & \textbf{Value}\\
 \bottomrule
Collection Size	& 6.7 MB\\
Document Format	& xml\\
Total Domains & 254\\
Total Queries	& 50\\
Total Token in Queries	& 219\\
Unique Tokens in Queries	& 143\\
Number of Categories	& 11\\
Total Documents	& 1,096\\
Average Documents for each Query	& 21\\
Total Token in Documents	& 859,432\\
Token after Stop Words Removal	& 461,300\\
Unique Tokens in Documents	& 15,418\\
Largest Document (by size)	& 143KB\\
Smallest Document (by size)	& 1KB\\
\bottomrule
\end{tabular}
\end{center}
\bigskip\centering
\end{minipage}
\end{table}

\subsection{Relevance judgments}
\label{subsec:Judgement}
After the creation of queries and pool of documents, relevance judgments are required for documents against the queries. For this purpose, we acquired assistance from the same subjects who created queries. The following guidelines were provided to subjects for relevance judgment:

\begin{itemize}
\item If a document is \textit{partially} relevant to a query, that document should be considered relevant. For example, in case of the third query of Fig.~\ref{fig:QIN}, (Pakistan dar-aamdat o bar-aamdat , Pakistani imports, and exports), all documents which provide information either related to ``Imports/exports"  or both will be considered as relevant to the query.

\item Documents in which the original query term does not exist, but a synonym of that word is used should be considered relevant as well. In the last query of Fig~\ref{fig:QIN}, a document is retrieved that does not contain the original word ``asbab (reason)" but a synonym ``waja (reason)" is used so that document is considered to be a relevant document for that query. 
\end{itemize}

Two of the three subjects provided binary (relevant, not relevant) judgments for all $1,096$ documents manually. Table.~\ref{tab:jud} shows the sample relevance judgment. The format first shows the query ID followed by the document ID and the relevance judgment for this query-document pair. In our notation, `1' value of relevance judgment means that the document is relevant to the query. After assigning relevance judgment to query-document pairs, the Kappa statistic was used to assess the degree of agreement between two subjects~\cite{cohen1960coefficient}.

\begin{table}
\caption{Sample relevance judgment from CURE}
\label{tab:jud}
\begin{minipage}{\columnwidth}
\begin{center}
\begin{tabular}{lll}
 \toprule
 \textbf{Query ID} & \textbf{Document ID} &\textbf{Judgment}\\
 \bottomrule
Q03 & CURE-0070 & 1\\
Q03 & CURE-0071 & 0\\
Q03 & CURE-0072 & 0\\
Q03 & CURE-0073 & 0\\
Q03	& CURE-0074 & 1\\
\bottomrule
\end{tabular}
\end{center}
\bigskip\centering
\end{minipage}
\end{table}

\begin{table}
\caption{Details of relevance judgment}
\label{tab:rel}
\begin{minipage}{\columnwidth}
\begin{center}
\begin{tabular}{llll}
 \toprule
 &\textbf{S1:Relevant} & \textbf{S1:Not-Relevant} & \textbf{Total}\\
\bottomrule
\textbf{S2:Relevant}	& 701	& 31 & 732\\
\textbf{S2:Not-Relevant} & 112 & 252 &364\\
\bottomrule
\textbf{Total}	& 813 & 283 &1096\\
\bottomrule
\end{tabular}
\end{center}
\bigskip\centering
\end{minipage}
\end{table}

\begin{equation}
\label{eqn:01}
Kappa=\frac{P(A)-P(E)}{1-P(E)}
\end{equation}

In the above equation~\ref{eqn:01}, P(A) is the proportion of time subjects agreed on relevance judgment while P(E) is the proportion of time subjects would be expected to agree by chance. From Table~\ref{tab:rel}, we found value of P(A) and P(E) is $0.87$ and $0.57$  respectively so `agreement score' is 70\%, which is considered to be a substantial agreement~\cite{gwet2012benchmarking}. In our study, subject-1 (S1) and subject-2 (S2) judged $813$ and $732$ documents to be relevant for 50 queries, respectively. Both subjects agreed on $701$ documents as relevant documents. A third subject (S3) was asked to resolve $143$ conflicting judgments to break the tie. From these $143$ documents, the third subject marked $92$ documents as relevant. Hence, from $1,096$ documents, $793$ documents were judged to be relevant documents against $50$ queries. After analyzing documents with conflicting relevance judgments, we observe two reasons that caused conflict in relevance judgments:

\begin{itemize}
\item Mostly subjects were confused in cases of lengthy queries. Therefore, we revised our guidelines and restricted query length to a maximum of $7$ words.
\item Subjects provided strict relevance judgment for their own created queries; however, their judgments were not strict for queries created by other subjects. This behavior was the main cause of conflicting relevance judgment.
\end{itemize}

\section{Methodology}
\label{sec:Methodology}
First, we explain the language preprocessing applied to each query. Next, we briefly discuss different retrieval models selected for the evaluation of our test collection. Finally, we will explain basic language processing techniques which we also evaluated on our test collection.

\subsection{Language pre-processing}
\label{subsec:Pre-Processing}
When the query is received, the first step is to preprocess the query. In the preprocessing phase, the query is tokenized based on white space delimiter. After tokenization, we remove any un-necessary punctuation marks. After this preprocessing, the query is issued to the retrieval models.

\subsection{Retrieval models}
\label{retrieval_model}
For evaluation, we have chosen four retrieval models: ``TF-IDF based VSM"~\cite{guo2008similarity}, ``BM-25 model"~\cite{robertson1995okapi}, ``LM-DS", and ``LM-JMS"~\cite{zhai2017study} for comparison. Before applying any retrieval model, we have applied OR-ed operation on queries with length greater than 1 to create the first subset of potentially relevant documents. After generation, the documents are ranked based on relevance scores provided by the retrieval models. Next, we provide a brief description of the selected IR models.

\begin{itemize}
\item \textbf{Term Frequency-Inverse Document Frequency (TF-IDF) based Vector Space Model:}
In this type of retrieval model, documents in the corpus and queries are transformed into vector space using a bag of words representation scheme. The weights used for the terms are TF-IDF weights. Term frequency gives us the number of occurrences of a term in a document, whereas IDF value gives the notion of the uniqueness of a term to a document. Each term in the index is converted into an n-dimensional vector where `n' is the number of documents. After transformation, \textit{cosine similarity} is computed between the query and the document. TF is calculated by taking the square root of the number of occurrences of a word in the document, and traditional IDF is modified to never output a zero value.  Cosine similarity and TF-IDF are calculated as follows~\cite{guo2008similarity}:

\begin{equation}
Cosine Similarity(q,d) = \frac{V_{q} . V_{d}}{|V_{q}|.|V_{d}|}
\end{equation}
Where Vq is the vector representation for query and Vd is the vector space representation for the document.

\begin{equation}
\small
TF-IDF (t,d)= (\sqrt{tf})*\left(1+log(\frac{ND}{DF+1})\right)
\end{equation}

Here, tf stands for the number of times a certain term `t' occurs in a document; ND is a number that represents total documents in the corpus and DF represents the document number that contains term `t'.
\newline
\item \textbf{Best Matching (BM25) Model:}
\newline
BM25 can be termed as the next generation of the TF-IDF vector space model. BM25 makes use of the saturation parameters to control the term frequency parameter. Moreover, the length of the document is incorporated into the final score calculation relative to the average of the length of all the documents in the corpus. Final score in the BM25 algorithm is calculated as follows~\cite{robertson1995okapi}:

\begin{equation}
\small
BM25(q,d)=\{\sum_{`t' in `q'}\left([log\left(\frac{\ ND-DF+0.5}{ DF+0.5}\right)]*[\frac{\ tf*(k1+1)}{ tf+k1(1-b+b * dL/avgdL)}]\right)\}
\end{equation}
Where dL is the document length and avgdL is the average document length in the corpus. $k1$ and $b$ are the saturation parameters and their values are $k1=1.2$ and $b=0.75$ in our setting. 
\newline
\item \textbf{Language Model with Dirichlet Smoothing (LM-DS):}
\newline
Language model for IR is a probabilistic distribution that is learned in the form of a query likelihood estimation for each document independently. Smoothing is used to adjust this estimation in case of a query term that is absent in a document. Note that, the smoothing parameter used in this model is \textit{dependent} on the document length. Using a constant \begin{math} \mu = 2000 \end{math} and the document length, the value for smoothing parameter \begin{math}(\lambda)\end{math} is determined as:

\begin{equation}
\lambda=1-\frac{N}{N+\mu}
\end{equation}

The overall probability with smoothing is calculated using the following\cite{zhai2017study}:

\begin{equation}
P_{\mu}(q|d)=\prod_{`t'in(q,d)}\left(\frac{c(t,d)+\mu.P(t|c)}{\sum_{t}c(t,d)+\mu}\right)
\end{equation}

where N is the document length and \begin{math}P_{ \mu}(q|d)\end{math} is the probability of the query being generated given a document. c(t,d) denotes the document frequency of a term. P(t|c) is the language model of the collection.
\newline
\item \textbf{Language Model with Jelenik Mercer Smoothing (LM-JMS):}
\newline
A linear interpolation of maximum likelihood estimation of a query word in the document and language model of a collection is employed in this model. The smoothing parameter is defined \textit{independent} of documents and the query and documents with higher lengths tend to provide better estimates. Optimal smoothing parameter value \begin{math} (\lambda) \end{math} needs to be tuned for the type of dataset under consideration.
In our setting, \begin{math} (\lambda) \end{math} is set to 0.7. The overall probability score for ranking is calculated as\cite{zhai2017study}: 

\begin{equation}
P_{\lambda}(q|d)=\prod_{`t'in(q,d)}\left((1-\lambda)P_{ml}(t|d)+(\lambda)P(t|c)\right)
\end{equation}
Here, \begin{math} P_{\lambda}(q|d) \end{math} is the probability of smoothing for a term in the document, \begin{math} P_{ml}(t|d) \end{math} is maximum likelihood estimation of term `t' in the document `d' whereas  \begin{math} P(t|c) \end{math} stands for the language model of the collection.
\end{itemize}

\subsection{Natural Language Processing techniques in IR}
Basic NLP techniques are essential requirements for any IR system. Our goal is to study the impact of enabling these NLP techniques on our test collection. Next, we describe different language processing techniques such as stop-words elimination, lemmatization, and query expansion that we used in addition to the baseline evaluation for the four retrieval models. 

\begin{figure}
\begin{center}
\includegraphics[width=8cm,height=3cm]{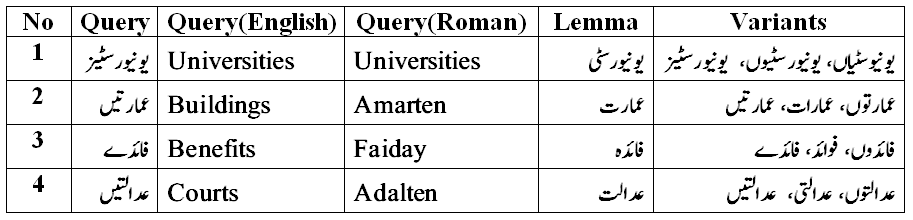}
\caption{Query words with lemma and variants}
\label{fig:var}
\end{center}
\end{figure}

\begin{itemize}
\item \textbf{Stop-words removal:}
Stop-words are eliminated to evaluate the performance of our test collection. Stop-words removal slightly improved the performance of IR on Kurdish test collection~\cite{esmaili2014towards}.
We removed stop-words from the query using a predefined Urdu stop-words list\footnote{\url{http://www.cle.org.pk/software/ling_resources/UrduClosedClassWordsList.htm}}. The stop-words list consists of very frequent words that do not carry any meaning of their own. There were $402$ words in this list, and we further reduced to $211$ words by removing cardinal and ordinal words.

\item \textbf{Lemmatization:}
Lemmatization is used to find the root or stem of the input word, and it is closely related to the stemmer. In~\cite{balakrishnan2014stemming}, authors have used stemming and lemmatization to improve precision in IR for the English language, and for evaluation, they used CACM collection. Mean average precision was used as evaluation metrics for top 10 and top 20 documents. They compared results of baseline technique with these NLP techniques and reported significant improvement. The main hurdle in the Urdu language is the scarcity of basic language modeling tools. Visali Gupta~\cite{gupta2016design} developed rule-based lemmatizer but it is not available for public use. Dictionary-based lemmatization with Stop-Words Removal (SWR) is applied at the query and index level, and we have $3,000$ word lemma pair in our lemmatization dictionary. In our test collection from $50$ queries, lemmatization affected only $24$ queries. Example of lemmatization is shown in Fig.~\ref{fig:var}.

\item \textbf{Query Expansion:}
Query expansion is a process of generating all inflectional forms of given input words in the query~\cite{carlberger2001improving}. Inflectional morpheme produces the grammatical form of a word (by adding plurals and a similar form of the word) and does not change part of speech~\cite{khan2011challenges}. In information retrieval, performance can be improved by expanding query words because when words in query are expanded, chances of mismatch between document and query are minimized~\cite{vechtomova2009query}. Effectiveness of information retrieval can also be improved by expanding the original query to synonyms~\cite{carpineto2012survey}. CURE is also used for the evaluation of query expansion with Stop-Words Removal (SWR). Most frequent $1000$ words were selected from our index to add inflectional variants, and after stop-words removal, we have $642$ unique words. We manually added inflectional forms of $642$ root words, and we have $2280$ variants for these words. The maximum number of variants of a word is $12$, and minimum $1$ while on average we have $3.55$ variants of a word. Example of variants is shown in Fig.~\ref{fig:var}. Our query word is first mapped to the root form of the word, and then variants of a word are added along with the root word. For the query ``Amarten (Buildings)'', three variants of root word ``Amart (Building)'' is added. We retrieved all documents in which ``Amarten (Buildings)'' or ``Amart (Building)'' or ``Amarton (Buildings)'' words are present.
\end{itemize}

\subsection{Ranking}
In general, we find the relevance score of the title and content field of the document for each word in the query for retrieval and ranking of test collection documents. The maximum of the relevance scores from title and content fields is taken as the relevance score of that query word against that document. While the relevance scores of individual query words are added in VSM and BM25 models, relevance scores are multiplied in case of language models to get the final score. Finally, retrieved results are sorted for the relevance score in descending order for ranking purpose.

\section{Results}
\label{sec:results}
In this section, we describe our experimental results for the evaluation of different information retrieval models based on our test collection. Furthermore, we also discuss the impact of different NLP techniques on the evaluation of these models. For this purpose, retrieved documents were compared against relevance judgment provided in test collection for different use cases. 

\subsection{Metrics definitions}
\label{subsec:metrics}
We use various different metrics for the evaluation such as \textit{Precision@10}, \textit{Precision@20}, \textit{Recall@50}, and \textit{Mean Average Precision (MAP)@50}. We briefly define these metrics below:

\begin{itemize}
\item Precision@k in information retrieval is the ratio of \textit{relevant results} in the top $k$ number of retrieved results. For instance, if five documents are relevant to a query in the top $10$ retrieved results, its precision will be $0.5$.

\item Recall@k is the ratio of relevant documents in the top $k$ retrieved results to the total number of relevant documents present in the corpus.

\item Mean Average Precision (MAP)@k is used to measure if the relevant results in top $k$ retrieved results are present in the top ranks. MAP is a single measure that spans mean of average precision over all queries.

\end{itemize}
 
\subsection{Evaluation of IR Models for CURE}
\label{subsec:Evaluation}
Table~\ref{tab:4th} shows Precision@$10$, Precision@$20$, Recall@$50$, and MAP for $50$ queries. In general, Precision@10 values for BM25 and LM with Jelenik Mercer smoothing for baseline perform in the range of $0.71-0.72$ whereas VSM the LM with Dirichlet smoothing shows a lower performance of $0.67-0.68$. A precision of $0.72$ means that on average $7$ out of $10$ retrieved documents are considered relevant. We observe a similar trend for Precision@20 for four retrieval model. We note that, as we increase the number of retrieved documents, precision will decrease while recall will increase~\cite{robertson2000evaluation}. Recall values in the top $50$ ranked documents are almost the same for every model except VSM with values above $0.85$ in every case which means over 85\% of the relevant documents against a query in the test collection are retrieved in the top $50$ retrieved results for each model. Our results are consistent with already reported results in~\cite{riaz2008baseline}. Similarly, MAP values for all the retrieval models under consideration are more than $0.67$. Such level of MAP value indicates that we will have more relevant results in the top ranks as compared to non-relevant results.
According to \cite{zhai2017study}, the performance of a particular retrieval model is highly dependent on the test collection under consideration and its attributes. A model performing well for a specific language test collection might not perform the same way on another data set. In our setting, we see that BM25 improves the retrieval results as compared to VSM. Also, we see a significant difference in the performance of language model retrievals where LM with Jelenik Mercer Smoothing outperforming LM with Dirichlet smoothing in every measure. This can be due to the nature of the language or the smoothing values used, but the exploration of this phenomena is left for future work.
\begin{table}
\caption{Results of retrieval models}
\label{tab:4th}
\begin{minipage}{\columnwidth}
\begin{center}
\begin{tabular}{llllll}
 \toprule
\textbf{Technique} & \textbf{Model} &\textbf{Precision@10} &\textbf{Precision@20} & \textbf{Recall@50} &\textbf{MAP@50}\\
\bottomrule
\multirow {4}{*}{Baseline} & BM25 & 0.73    & 0.62    & 0.97    & 0.73\\
 & VSM    & 0.66    & 0.57    & 0.86    & 0.69\\
 & LM-DS     & 0.69    & 0.58    & 0.93    & 0.68\\
 & LM-JMS    & 0.73    & 0.61    & 0.97    & 0.73\\
\hline
\multirow {4}{*}{Stop-Words-Removal (SWR)} & BM25     & 0.73    & 0.62    & 0.97    & 0.74\\
    & VSM  & 0.69    & 0.57    & 0.86    & 0.70\\
    & LM-DS & 0.69    & 0.58    & 0.93    & 0.68\\
    & LM-JMS & 0.73    & 0.61    & 0.96    & 0.73\\
\hline
\multirow {4}{*}{Lemmatization with SWR} & BM25 & 0.74    & 0.63    & 0.97    & 0.75\\
& VSM & 0.71    & 0.58    & 0.87    & 0.72\\
& LM-DS    & 0.71    & 0.59    & 0.93    & 0.70\\
& LM-JMS & 0.75    & 0.63    & 0.97    & 0.75\\
\hline
\multirow {4}{*}{Query-Expansion with SWR} & BM25     & 0.74    & 0.63    & 0.98    & 0.75\\
& VSM & 0.71    & 0.58    & 0.88    & 0.72\\
 & LM-DS & 0.70    & 0.58    & 0.94    & 0.69\\
 & LM-JMS    & 0.75    & 0.63    & 0.98    & 0.75\\
 \bottomrule
\end{tabular}
\end{center}
\bigskip\centering
\end{minipage}
\end{table}

\subsection{Error Analysis for Each Query}
\label{subsec:errorAnalysis}
In our study, LM-JMS (P@10) returned at least nine correct relevant documents for 30 out of 50 total queries. Our investigation of the remaining queries showed that irrelevant documents were retrieved due to the retrieval model and errors in NLP techniques for 11 and 9 queries, respectively. Keeping in view the proposed NLP tools pipeline for our experiments in indexing and retrieving, we carried out detailed error analysis for 20 queries. The query-by-query error analysis of queries can be downloaded from \footnote{\url{https://www.kics.edu.pk/kics_pms/documents/projects/deliverables/deliverable_357_Query_by_query_error_analysis.pdf}}.

For a given query, the order of NLP tools in our IR system pipeline is as follows: 1) word tokenization, 2) stop words removal, 3) lemmatization, and 4) query expansion. Our query by query error analysis showed that overall retrieval performance dropped due to an error in various NLP tools. Our performance was decreased due to word tokenizer (2 queries), lemmatization (2 queries), and query expansion (5 queries). Below we discuss detailed error analysis for each NLP tool that resulted in a decrease of overall retrieval performance.

\begin{figure}
\begin{center}
\includegraphics[width=10cm,height=10cm]{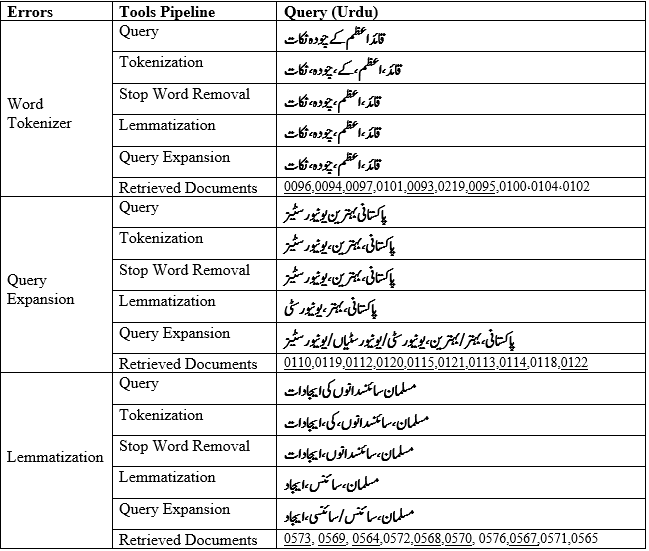}
\caption{Error Analysis of NLP Techniques}
\label{fig:nlpAnalysisiError}
\end{center}
\end{figure}

Fig~\ref{fig:nlpAnalysisiError} shows an example of errors introduced by different NLP techniques applied in the IR tools pipeline. We start with an example of errors based on word tokenizer. First of all, our query was tokenized, and five tokens were obtained. However, four tokens were returned by stop-word removal, lemmatization, and query expansion. Note that we used lookup based lemmatization and query expansion in such a way that our dictionary did not cover the root word and variants for query terms. In the sample shown, `Quaid Azam" is one term; however, our word tokenizer tokenized this query based on white space in two tokens, i.e., ``Quaid" and ``Azam". For this query, a total of five relevant documents were retrieved underlined in the figure~\ref{fig:nlpAnalysisiError}. Some relevant documents were not retrieved for this query because of two reasons: 1) keyword ``Jinnah" (actual name) was present in the documents instead of ``Quaid Azam", and 2) there was no space present between the term ``Quaid" and ``Azam".

Next, we discuss the impact of query expansion. For this query, a total of eight relevant documents were retrieved on the top ten ranks. Our detailed inspection of both irrelevant documents showed that keywords ``jamia (institute)", ``jamiat (institutes)", and ``institutes" were used instead of ``university". Error in this query occurred mainly due to query expansion because we do not have these synonyms for the query expansion. Similarly, a total of six relevant documents and four irrelevant documents were retrieved on the top ten ranks in case of lemmatization. Here the root word ``Science" provided by the lemmatizer was matched with the index term present in retrieved documents. Due to this reason, less relevant documents were pushed to higher ranks after lemmatization for this query.

\begin{figure}
\begin{center}
\includegraphics[width=8cm,height=4cm]{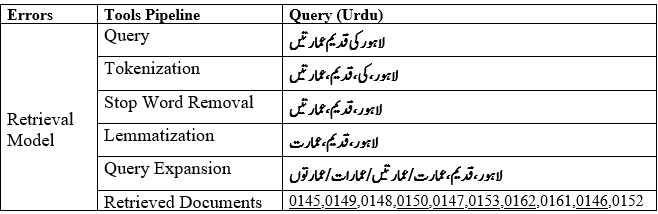}
\caption{Error Analysis of LM-JMS}
\label{fig:retModel}
\end{center}
\end{figure}

Other than NLP techniques, our retrieval models also triggered some errors. Fig~\ref{fig:retModel} shows an example of error occurred due to the retrieval model. For the last query mostly those documents were retrieved in which the terms ``qadeem (ancient)" or ``dunia ke qadeem amarteen (the ancient buildings of the world)" occurred frequently.

In general, for all retrieval models, mostly those documents were retrieved that include original query words for the baseline case. However, for lemmatization and query expansion retrieval of top rank documents relied on root words and variants of query terms. In our case, most errors were observed for query expansion. We believe this performance can further be enhanced by adding synonyms in the dictionary of query expansion. By adding synonyms, chances of vocabulary mismatch will be decreased for efficient retrieval. Our evaluation results indicate that by applying lemmatization and query expansion, results can be improved for Urdu IR which is in accordance with ~\cite{balakrishnan2014stemming} where authors use language pre-processing techniques such as stemming and lemmatization at document level on the English IR to improve retrieval performance.

{
}

\section{Conclusion}
\label{sec:conclusion}

Standardized evaluation of an IR system cannot be achieved in the absence of a test collection. The availability of such resources in the Urdu language is very scarce.  In this research work, we have developed a standard test collection (CURE) for the task of Urdu IR evaluation. We manually designed $50$ queries and top $20$ documents were retrieved using VSM and BM25 models from a collection of 0.5 million documents for each query. After removing $904$ duplicate documents, we have $1096$ unique documents in our test collection. For each document, binary relevance judgment is provided. Besides, we have developed essential NLP resources such as stop-words removal, lemmatization, and query expansion that are essential for information retrieval.  

To evaluate the performance of stop-words elimination, already developed list was used. For evaluation of lemmatization and query expansion, we manually developed resource from most frequent words of indexed documents. In lemmatization dictionary, we have root form of $3000$ words and in query expansion $2280$ inflectional forms of $642$ words. Three evaluation measures  Precision, Recall, and MAP was used to evaluate the performance of baseline retrieval models in addition to the evaluation of language resources. Results show that the performance of all retrieval model was improved by using basic language processing techniques. This test collection can be used to evaluate the performance of different features of information retrieval systems and the application of different NLP techniques in IR. In the future, we plan to increase the number of documents and queries in our collection. Complete test collection contains a set of queries, document collection, relevance judgment, and list of stop words \footnote{\url{http://kics.edu.pk/kics_pms/index.php?r=user/auth}}.


\renewcommand{\shortauthors}{M. Iqbal et al.}

\bibliographystyle{ACM-Reference-Format}
\bibliography{main}


\begin{thebibliography}{31}


\ifx \showCODEN    \undefined \def \showCODEN     #1{\unskip}     \fi
\ifx \showDOI      \undefined \def \showDOI       #1{#1}\fi
\ifx \showISBNx    \undefined \def \showISBNx     #1{\unskip}     \fi
\ifx \showISBNxiii \undefined \def \showISBNxiii  #1{\unskip}     \fi
\ifx \showISSN     \undefined \def \showISSN      #1{\unskip}     \fi
\ifx \showLCCN     \undefined \def \showLCCN      #1{\unskip}     \fi
\ifx \shownote     \undefined \def \shownote      #1{#1}          \fi
\ifx \showarticletitle \undefined \def \showarticletitle #1{#1}   \fi
\ifx \showURL      \undefined \def \showURL       {\relax}        \fi
\providecommand\bibfield[2]{#2}
\providecommand\bibinfo[2]{#2}
\providecommand\natexlab[1]{#1}
\providecommand\showeprint[2][]{arXiv:#2}

\bibitem[\protect\citeauthoryear{AleAhmad, Amiri, Darrudi, Rahgozar, and
  Oroumchian}{AleAhmad et~al\mbox{.}}{2009}]%
        {aleahmad2009hamshahri}
\bibfield{author}{\bibinfo{person}{Abolfazl AleAhmad}, \bibinfo{person}{Hadi
  Amiri}, \bibinfo{person}{Ehsan Darrudi}, \bibinfo{person}{Masoud Rahgozar},
  {and} \bibinfo{person}{Farhad Oroumchian}.} \bibinfo{year}{2009}\natexlab{}.
\newblock \showarticletitle{Hamshahri: A standard Persian text collection}.
\newblock \bibinfo{journal}{\emph{Knowledge-Based Systems}}
  \bibinfo{volume}{22}, \bibinfo{number}{5} (\bibinfo{year}{2009}),
  \bibinfo{pages}{382--387}.
\newblock


\bibitem[\protect\citeauthoryear{Balakrishnan and Lloyd-Yemoh}{Balakrishnan and
  Lloyd-Yemoh}{2014}]%
        {balakrishnan2014stemming}
\bibfield{author}{\bibinfo{person}{Vimala Balakrishnan} {and}
  \bibinfo{person}{Ethel Lloyd-Yemoh}.} \bibinfo{year}{2014}\natexlab{}.
\newblock \showarticletitle{Stemming and lemmatization: a comparison of
  retrieval performances}.
\newblock \bibinfo{journal}{\emph{Lecture Notes on Software Engineering}}
  \bibinfo{volume}{2}, \bibinfo{number}{3} (\bibinfo{year}{2014}),
  \bibinfo{pages}{262}.
\newblock


\bibitem[\protect\citeauthoryear{Becker and Riaz}{Becker and Riaz}{2002}]%
        {becker2002study}
\bibfield{author}{\bibinfo{person}{Dara Becker} {and} \bibinfo{person}{Kashif
  Riaz}.} \bibinfo{year}{2002}\natexlab{}.
\newblock \showarticletitle{A study in urdu corpus construction}. In
  \bibinfo{booktitle}{\emph{Proceedings of the 3rd workshop on Asian language
  resources and international standardization-Volume 12}}. Association for
  Computational Linguistics, \bibinfo{pages}{1--5}.
\newblock


\bibitem[\protect\citeauthoryear{Carlberger, Dalianis, Duneld, and
  Knutsson}{Carlberger et~al\mbox{.}}{2001}]%
        {carlberger2001improving}
\bibfield{author}{\bibinfo{person}{Johan Carlberger}, \bibinfo{person}{Hercules
  Dalianis}, \bibinfo{person}{Martin Duneld}, {and} \bibinfo{person}{Ola
  Knutsson}.} \bibinfo{year}{2001}\natexlab{}.
\newblock \showarticletitle{Improving precision in information retrieval for
  Swedish using stemming}. In \bibinfo{booktitle}{\emph{Proceedings of the 13th
  Nordic Conference of Computational Linguistics (NODALIDA 2001)}}.
\newblock


\bibitem[\protect\citeauthoryear{Carpineto and Romano}{Carpineto and
  Romano}{2012}]%
        {carpineto2012survey}
\bibfield{author}{\bibinfo{person}{Claudio Carpineto} {and}
  \bibinfo{person}{Giovanni Romano}.} \bibinfo{year}{2012}\natexlab{}.
\newblock \showarticletitle{A survey of automatic query expansion in
  information retrieval}.
\newblock \bibinfo{journal}{\emph{ACM Computing Surveys (CSUR)}}
  \bibinfo{volume}{44}, \bibinfo{number}{1} (\bibinfo{year}{2012}),
  \bibinfo{pages}{1}.
\newblock


\bibitem[\protect\citeauthoryear{Clough and Sanderson}{Clough and
  Sanderson}{2013}]%
        {Paul2013Evaluating}
\bibfield{author}{\bibinfo{person}{Paul Clough} {and} \bibinfo{person}{Mark
  Sanderson}.} \bibinfo{year}{2013}\natexlab{}.
\newblock \showarticletitle{Evaluating the performance of information retrieval
  systems using test collections.}
\newblock \bibinfo{journal}{\emph{Information Research}} \bibinfo{volume}{18},
  \bibinfo{number}{2} (\bibinfo{year}{2013}).
\newblock


\bibitem[\protect\citeauthoryear{Cohen}{Cohen}{1960}]%
        {cohen1960coefficient}
\bibfield{author}{\bibinfo{person}{Jacob Cohen}.}
  \bibinfo{year}{1960}\natexlab{}.
\newblock \showarticletitle{A coefficient of agreement for nominal scales}.
\newblock \bibinfo{journal}{\emph{Educational and psychological measurement}}
  \bibinfo{volume}{20}, \bibinfo{number}{1} (\bibinfo{year}{1960}),
  \bibinfo{pages}{37--46}.
\newblock


\bibitem[\protect\citeauthoryear{Esmaili, Abolhassani, Neshati, Behrangi,
  Rostami, and Nasiri}{Esmaili et~al\mbox{.}}{2007}]%
        {esmaili2007mahak}
\bibfield{author}{\bibinfo{person}{Kyumars~Sheykh Esmaili},
  \bibinfo{person}{Hassan Abolhassani}, \bibinfo{person}{Mahmood Neshati},
  \bibinfo{person}{Ehsan Behrangi}, \bibinfo{person}{Asreen Rostami}, {and}
  \bibinfo{person}{Mojtaba~Mohammadi Nasiri}.} \bibinfo{year}{2007}\natexlab{}.
\newblock \showarticletitle{Mahak: A test collection for evaluation of Farsi
  information retrieval systems}. In \bibinfo{booktitle}{\emph{Computer Systems
  and Applications, 2007. AICCSA'07. IEEE/ACS International Conference on}}.
  IEEE, \bibinfo{pages}{639--644}.
\newblock


\bibitem[\protect\citeauthoryear{Esmaili, Salavati, and Datta}{Esmaili
  et~al\mbox{.}}{2014}]%
        {esmaili2014towards}
\bibfield{author}{\bibinfo{person}{Kyumars~Sheykh Esmaili},
  \bibinfo{person}{Shahin Salavati}, {and} \bibinfo{person}{Anwitaman Datta}.}
  \bibinfo{year}{2014}\natexlab{}.
\newblock \showarticletitle{Towards kurdish information retrieval}.
\newblock \bibinfo{journal}{\emph{ACM Transactions on Asian Language
  Information Processing (TALIP)}} \bibinfo{volume}{13}, \bibinfo{number}{2}
  (\bibinfo{year}{2014}), \bibinfo{pages}{7}.
\newblock


\bibitem[\protect\citeauthoryear{Fox}{Fox}{1983}]%
        {fox1983characterization}
\bibfield{author}{\bibinfo{person}{Edward~A Fox}.}
  \bibinfo{year}{1983}\natexlab{}.
\newblock \bibinfo{booktitle}{\emph{Characterization of two new experimental
  collections in computer and information science containing textual and
  bibliographic concepts}}.
\newblock \bibinfo{type}{{T}echnical {R}eport}. \bibinfo{institution}{Cornell
  University}.
\newblock


\bibitem[\protect\citeauthoryear{Guo}{Guo}{2008}]%
        {guo2008similarity}
\bibfield{author}{\bibinfo{person}{Qinglin Guo}.}
  \bibinfo{year}{2008}\natexlab{}.
\newblock \showarticletitle{The similarity computing of documents based on
  VSM}. In \bibinfo{booktitle}{\emph{International Conference on Network-Based
  Information Systems}}. Springer, \bibinfo{pages}{142--148}.
\newblock


\bibitem[\protect\citeauthoryear{Gupta, Joshi, and Mathur}{Gupta
  et~al\mbox{.}}{2016}]%
        {gupta2016design}
\bibfield{author}{\bibinfo{person}{Vaishali Gupta}, \bibinfo{person}{Nisheeth
  Joshi}, {and} \bibinfo{person}{Iti Mathur}.} \bibinfo{year}{2016}\natexlab{}.
\newblock \showarticletitle{Design and development of a rule-based Urdu
  lemmatizer}. In \bibinfo{booktitle}{\emph{Proceedings of International
  Conference on ICT for Sustainable Development}}. Springer,
  \bibinfo{pages}{161--169}.
\newblock


\bibitem[\protect\citeauthoryear{Gwet}{Gwet}{2012}]%
        {gwet2012benchmarking}
\bibfield{author}{\bibinfo{person}{KL Gwet}.} \bibinfo{year}{2012}\natexlab{}.
\newblock \showarticletitle{Benchmarking inter-rater reliability coefficients}.
\newblock \bibinfo{journal}{\emph{Handbook of inter-rater reliability. 3rd edn.
  Gaithersburg, MD}} (\bibinfo{year}{2012}), \bibinfo{pages}{121--128}.
\newblock


\bibitem[\protect\citeauthoryear{Hardie}{Hardie}{2003}]%
        {hardie2003developing}
\bibfield{author}{\bibinfo{person}{Andrew Hardie}.}
  \bibinfo{year}{2003}\natexlab{}.
\newblock \showarticletitle{Developing a tagset for automated part-of-speech
  tagging in Urdu.}. In \bibinfo{booktitle}{\emph{Corpus Linguistics 2003}}.
\newblock


\bibitem[\protect\citeauthoryear{Harman}{Harman}{1995}]%
        {harman1995overview}
\bibfield{author}{\bibinfo{person}{Donna Harman}.}
  \bibinfo{year}{1995}\natexlab{}.
\newblock \showarticletitle{Overview of the fourth text retrieval conference
  (TREC-4)}. In \bibinfo{booktitle}{\emph{The Fourth Text REtrieval Conference
  (TREC-4)}}. \bibinfo{pages}{1--24}.
\newblock


\bibitem[\protect\citeauthoryear{Hladek, Stas, and Juhar}{Hladek
  et~al\mbox{.}}{2014}]%
        {hladek2014slovak}
\bibfield{author}{\bibinfo{person}{Daniel Hladek}, \bibinfo{person}{Jan Stas},
  {and} \bibinfo{person}{Jozef Juhar}.} \bibinfo{year}{2014}\natexlab{}.
\newblock \showarticletitle{The Slovak Categorized News Corpus.}. In
  \bibinfo{booktitle}{\emph{LREC}}. \bibinfo{pages}{1705--1708}.
\newblock


\bibitem[\protect\citeauthoryear{Hl{\'a}dek, Stas, and Juh{\'a}r}{Hl{\'a}dek
  et~al\mbox{.}}{2016}]%
        {hladek2016evaluation}
\bibfield{author}{\bibinfo{person}{Daniel Hl{\'a}dek}, \bibinfo{person}{J{\'a}n
  Stas}, {and} \bibinfo{person}{Jozef Juh{\'a}r}.}
  \bibinfo{year}{2016}\natexlab{}.
\newblock \showarticletitle{Evaluation Set for Slovak News Information
  Retrieval.}. In \bibinfo{booktitle}{\emph{LREC}}.
\newblock


\bibitem[\protect\citeauthoryear{Khan, Anwar, and Bajwa}{Khan
  et~al\mbox{.}}{2011}]%
        {khan2011challenges}
\bibfield{author}{\bibinfo{person}{Sajjad~Ahmad Khan}, \bibinfo{person}{Waqas
  Anwar}, {and} \bibinfo{person}{Usama~Ijaz Bajwa}.}
  \bibinfo{year}{2011}\natexlab{}.
\newblock \showarticletitle{Challenges in developing a rule based urdu
  stemmer}. In \bibinfo{booktitle}{\emph{Proceedings of the 2nd Workshop on
  South Southeast Asian Natural Language Processing (WSSANLP)}}.
  \bibinfo{pages}{46--51}.
\newblock


\bibitem[\protect\citeauthoryear{Kinney, Huffman, and Zhai}{Kinney
  et~al\mbox{.}}{2008}]%
        {kinney2008evaluator}
\bibfield{author}{\bibinfo{person}{Kenneth~A Kinney}, \bibinfo{person}{Scott~B
  Huffman}, {and} \bibinfo{person}{Juting Zhai}.}
  \bibinfo{year}{2008}\natexlab{}.
\newblock \showarticletitle{How evaluator domain expertise affects search
  result relevance judgments}. In \bibinfo{booktitle}{\emph{Proceedings of the
  17th ACM conference on Information and knowledge management}}. ACM,
  \bibinfo{pages}{591--598}.
\newblock


\bibitem[\protect\citeauthoryear{Madankar, Chandak, and Chavhan}{Madankar
  et~al\mbox{.}}{2016}]%
        {madankar2016information}
\bibfield{author}{\bibinfo{person}{Mangala Madankar}, \bibinfo{person}{MB
  Chandak}, {and} \bibinfo{person}{Nekita Chavhan}.}
  \bibinfo{year}{2016}\natexlab{}.
\newblock \showarticletitle{Information retrieval system and machine
  translation: a review}.
\newblock \bibinfo{journal}{\emph{Procedia Computer Science}}
  \bibinfo{volume}{78} (\bibinfo{year}{2016}), \bibinfo{pages}{845--850}.
\newblock


\bibitem[\protect\citeauthoryear{Manning, Raghavan, Sch{\"u}tze,
  et~al\mbox{.}}{Manning et~al\mbox{.}}{2008}]%
        {manning2008introduction}
\bibfield{author}{\bibinfo{person}{Christopher~D Manning},
  \bibinfo{person}{Prabhakar Raghavan}, \bibinfo{person}{Hinrich Sch{\"u}tze},
  {et~al\mbox{.}}} \bibinfo{year}{2008}\natexlab{}.
\newblock \bibinfo{booktitle}{\emph{Introduction to information retrieval}}.
  Vol.~\bibinfo{volume}{1}.
\newblock \bibinfo{publisher}{Cambridge university press Cambridge}.
\newblock


\bibitem[\protect\citeauthoryear{Riaz}{Riaz}{2008}]%
        {riaz2008baseline}
\bibfield{author}{\bibinfo{person}{Kashif Riaz}.}
  \bibinfo{year}{2008}\natexlab{}.
\newblock \showarticletitle{Baseline for Urdu IR evaluation}. In
  \bibinfo{booktitle}{\emph{Proceedings of the 2nd ACM workshop on Improving
  non english web searching}}. ACM, \bibinfo{pages}{97--100}.
\newblock


\bibitem[\protect\citeauthoryear{Robertson}{Robertson}{2000}]%
        {robertson2000evaluation}
\bibfield{author}{\bibinfo{person}{Stephen Robertson}.}
  \bibinfo{year}{2000}\natexlab{}.
\newblock \showarticletitle{Evaluation in information retrieval}.
\newblock In \bibinfo{booktitle}{\emph{Lectures on information retrieval}}.
  \bibinfo{publisher}{Springer}, \bibinfo{pages}{81--92}.
\newblock


\bibitem[\protect\citeauthoryear{Robertson, Walker, Jones, Hancock-Beaulieu,
  Gatford, et~al\mbox{.}}{Robertson et~al\mbox{.}}{1995}]%
        {robertson1995okapi}
\bibfield{author}{\bibinfo{person}{Stephen~E Robertson}, \bibinfo{person}{Steve
  Walker}, \bibinfo{person}{Susan Jones}, \bibinfo{person}{Micheline~M
  Hancock-Beaulieu}, \bibinfo{person}{Mike Gatford}, {et~al\mbox{.}}}
  \bibinfo{year}{1995}\natexlab{}.
\newblock \showarticletitle{Okapi at TREC-3}.
\newblock \bibinfo{journal}{\emph{Nist Special Publication Sp}}
  \bibinfo{volume}{109} (\bibinfo{year}{1995}), \bibinfo{pages}{109}.
\newblock


\bibitem[\protect\citeauthoryear{Sakai, Kitani, Ogawa, Ishikawa, Kimoto, Keshi,
  Toyoura, Fukushima, Matsui, Ueda, et~al\mbox{.}}{Sakai et~al\mbox{.}}{1999}]%
        {sakai1999bmir}
\bibfield{author}{\bibinfo{person}{Tetsuya Sakai}, \bibinfo{person}{Tsuyoshi
  Kitani}, \bibinfo{person}{Yasushi Ogawa}, \bibinfo{person}{Tetsuya Ishikawa},
  \bibinfo{person}{Haruo Kimoto}, \bibinfo{person}{Ikuo Keshi},
  \bibinfo{person}{Jun Toyoura}, \bibinfo{person}{Toshikazu Fukushima},
  \bibinfo{person}{Kunio Matsui}, \bibinfo{person}{Yoshihiro Ueda},
  {et~al\mbox{.}}} \bibinfo{year}{1999}\natexlab{}.
\newblock \showarticletitle{BMIR-J2: a test collection for evaluation of
  Japanese information retrieval systems}. In \bibinfo{booktitle}{\emph{ACM
  SIGIR Forum}}, Vol.~\bibinfo{volume}{33}. ACM, \bibinfo{pages}{13--17}.
\newblock


\bibitem[\protect\citeauthoryear{Salton}{Salton}{1973}]%
        {salton1973recent}
\bibfield{author}{\bibinfo{person}{Gerard Salton}.}
  \bibinfo{year}{1973}\natexlab{}.
\newblock \showarticletitle{Recent studies in automatic text analysis and
  document retrieval}.
\newblock \bibinfo{journal}{\emph{Journal of the ACM (JACM)}}
  \bibinfo{volume}{20}, \bibinfo{number}{2} (\bibinfo{year}{1973}),
  \bibinfo{pages}{258--278}.
\newblock


\bibitem[\protect\citeauthoryear{Shamshed and Karim}{Shamshed and
  Karim}{2010}]%
        {shamshed2010novel}
\bibfield{author}{\bibinfo{person}{Jubayer Shamshed} {and}
  \bibinfo{person}{SM~Masud Karim}.} \bibinfo{year}{2010}\natexlab{}.
\newblock \showarticletitle{A Novel Bangla text corpus building method for
  efficient information retrieval}.
\newblock \bibinfo{journal}{\emph{Journal of Convergence Information
  Technology}} \bibinfo{volume}{1}, \bibinfo{number}{1} (\bibinfo{year}{2010}),
  \bibinfo{pages}{36--40}.
\newblock


\bibitem[\protect\citeauthoryear{Urbano}{Urbano}{2016}]%
        {urbano2016test}
\bibfield{author}{\bibinfo{person}{Juli{\'a}n Urbano}.}
  \bibinfo{year}{2016}\natexlab{}.
\newblock \showarticletitle{Test collection reliability: a study of bias and
  robustness to statistical assumptions via stochastic simulation}.
\newblock \bibinfo{journal}{\emph{Information Retrieval Journal}}
  \bibinfo{volume}{19}, \bibinfo{number}{3} (\bibinfo{year}{2016}),
  \bibinfo{pages}{313--350}.
\newblock


\bibitem[\protect\citeauthoryear{Vechtomova}{Vechtomova}{2009}]%
        {vechtomova2009query}
\bibfield{author}{\bibinfo{person}{Olga Vechtomova}.}
  \bibinfo{year}{2009}\natexlab{}.
\newblock \showarticletitle{Query expansion for information retrieval}.
\newblock In \bibinfo{booktitle}{\emph{Encyclopedia of database systems}}.
  \bibinfo{publisher}{Springer}, \bibinfo{pages}{2254--2257}.
\newblock


\bibitem[\protect\citeauthoryear{Weber}{Weber}{2008}]%
        {weber2008top}
\bibfield{author}{\bibinfo{person}{George Weber}.}
  \bibinfo{year}{2008}\natexlab{}.
\newblock \showarticletitle{Top languages}.
\newblock \bibinfo{journal}{\emph{The World's}}  \bibinfo{volume}{10}
  (\bibinfo{year}{2008}).
\newblock


\bibitem[\protect\citeauthoryear{Zhai and Lafferty}{Zhai and Lafferty}{2017}]%
        {zhai2017study}
\bibfield{author}{\bibinfo{person}{Chengxiang Zhai} {and} \bibinfo{person}{John
  Lafferty}.} \bibinfo{year}{2017}\natexlab{}.
\newblock \showarticletitle{A study of smoothing methods for language models
  applied to ad hoc information retrieval}. In \bibinfo{booktitle}{\emph{ACM
  SIGIR Forum}}, Vol.~\bibinfo{volume}{51}. ACM, \bibinfo{pages}{268--276}.
\newblock


\end{thebibliography}

\end{document}